\documentclass[aps,amsmath,preprint,epsf,superscriptaddress,nofootinbib]{revtex4-1}
\usepackage{graphicx}
\usepackage{bm}
\usepackage{color}
\usepackage{amsmath,amssymb}	
\usepackage{hyperref}
\usepackage{setspace}
\usepackage{longtable}
\usepackage{booktabs}

\pdfoutput=1

\def\slashchar#1{\setbox0=\hbox{$#1$} 
\dimen0=\wd0 
\setbox1=\hbox{/} \dimen1=\wd1 
\ifdim\dimen0>\dimen1 
\rlap{\hbox to \dimen0{\hfil/\hfil}} 
#1 
\else 
\rlap{\hbox to \dimen1{\hfil$#1$\hfil}} 
/ 
\fi}

\begin{document}

\preprint{IPMU16-0178}
\bigskip

\title{Constraints on $L_\mu-L_\tau$ Gauge Interactions from Rare Kaon Decay}

\author{Masahiro Ibe}
\email[e-mail: ]{ibe@icrr.u-tokyo.ac.jp}
\affiliation{Kavli IPMU (WPI), UTIAS, The University of Tokyo, Kashiwa, Chiba 277-8583, Japan}
\affiliation{ICRR, The University of Tokyo, Kashiwa, Chiba 277-8582, Japan}
\author{Wakutaka Nakano}
\email[e-mail: ]{m156077@icrr.u-tokyo.ac.jp}
\affiliation{Kavli IPMU (WPI), UTIAS, The University of Tokyo, Kashiwa, Chiba 277-8583, Japan}
\affiliation{ICRR, The University of Tokyo, Kashiwa, Chiba 277-8582, Japan}
\author{Motoo Suzuki}
\email[e-mail: ]{m0t@icrr.u-tokyo.ac.jp}
\affiliation{Kavli IPMU (WPI), UTIAS, The University of Tokyo, Kashiwa, Chiba 277-8583, Japan}
\affiliation{ICRR, The University of Tokyo, Kashiwa, Chiba 277-8582, Japan}

\date{\today}

\begin{abstract}
A model with  $L_\mu -L_\tau$ gauge symmetry is the least constrained model 
as a resolution to the disagreement of the muon anomalous magnetic moment between the 
theoretical predictions and the experimental results.
In this paper, we discuss how well the  $L_\mu - L_\tau$  model can be constrained 
by looking for decays of the charged kaon associated with
 a $L_\mu-L_\tau$ gauge boson.
More concretely, we consider  searches for single muon tracks from the decays of  stopped charged kaon
as in the E949 experiment.
In our conservative estimation, we find that the favored parameter region for the muon anomalous magnetic moment can be 
tested by using a $10$ times larger number of the stopped charged kaons and about a $100$ times better photon rejection rate
than the E949 experiment. 
\end{abstract}

\maketitle

\section{Introduction}
The Standard Model (SM) of particle physics has passed high precision experimental tests 
for decades as the best description of electroweak and strong interactions.
Despite this overwhelming success, however, 
there is a long-standing discrepancy between the SM 
prediction and the experimental value of the muon anomalous magnetic moment, $a_\mu = (g-2)_\mu/2$.
The world average of the measurements of $a_\mu$
is given by
\begin{eqnarray}
a_\mu = 116592091(63)\times 10^{-11}\ ,
\end{eqnarray}
which is dominated by the E821 experiment at the Brookhaven National 
Laboratory~\cite{Bennett:2006fi,Mohr:2012tt,Agashe:2014kda,Nyffeler:2016gnb}.
The SM prediction has comparable accuracy which includes QED corrections to five loops, 
weak corrections to two loops, and hadronic corrections~\cite[see][for review]{Jegerlehner:2009ry,Blum:2013xva,Benayoun:2014tra,Knecht:2014sea}.
Currently, there is a stable $3$--$4\,\sigma$ deviation between the measured and the predicted values,
where the range of the discrepancy is due to the uncertainties in the hadronic corrections.
Future experiments, such as E34~\cite{E34} (an ultra-cold muon beam experiment) and E989~\cite{Grange:2015fou} 
(an improved version of E821), are expected to reduce  experimental uncertainties by a factor of four 
which would confirm the discrepancy at the $5\sigma$ level.
In the following, we use the deviation between the experiment and the SM prediction in \cite{Olive:2016xmw}.
\begin{eqnarray}
\delta a_\mu = a_\mu^{\rm exp} - a_\mu^{\rm SM} = \left(280\pm 80\right)\times 10^{-11}\ .
\end{eqnarray}

So far, many SM extensions have been proposed to resolve the discrepancy~\cite[see][for review]{Jegerlehner:2009ry}.
Among various proposals, a class of models with a $U(1)$ gauge symmetry provides one of the most minimal extensions.
There,  the muon anomalous magnetic moment receives a contribution from the  massive gauge boson $Z'$ at the one-loop level,
which reduces the discrepancy.
The models with a $U(1)$ gauge symmetry are, however, severely constrained from numerous experimental searches.
For example, when the $U(1)$ gauge symmetry is identified with the $B-L$  gauge symmetry, 
the parameter region which resolves the discrepancy is excluded by neutrino-electron scattering 
experiments~\cite{Harnik:2012ni,Bilmis:2015lja}. 
Models with a dark photon are also studied extensively, where the dark photon has a kinetic mixing with the SM  photon.
However, those models are also excluded as a solution to the discrepancy by electron beam dump 
experiments~\cite{Andreas:2012mt,Babusci:2012cr,Adlarson:2013eza,Agakishiev:2013fwl,Merkel:2014avp,Merkel:2014avp,Batley:2015lha} 
and by $e^+e^-$ collider experiments~\cite{Lees:2014xha}.%
\footnote{Some of the constraints can be relaxed when the dark photon 
decays into a light dark matter (see e.g. Refs.~\cite{Davoudiasl:2014kua,Harigaya:2016rwr}).
}

\begin{figure}[t]
\begin{center}
\begin{minipage}{.5\linewidth}
  \includegraphics[width=\linewidth]{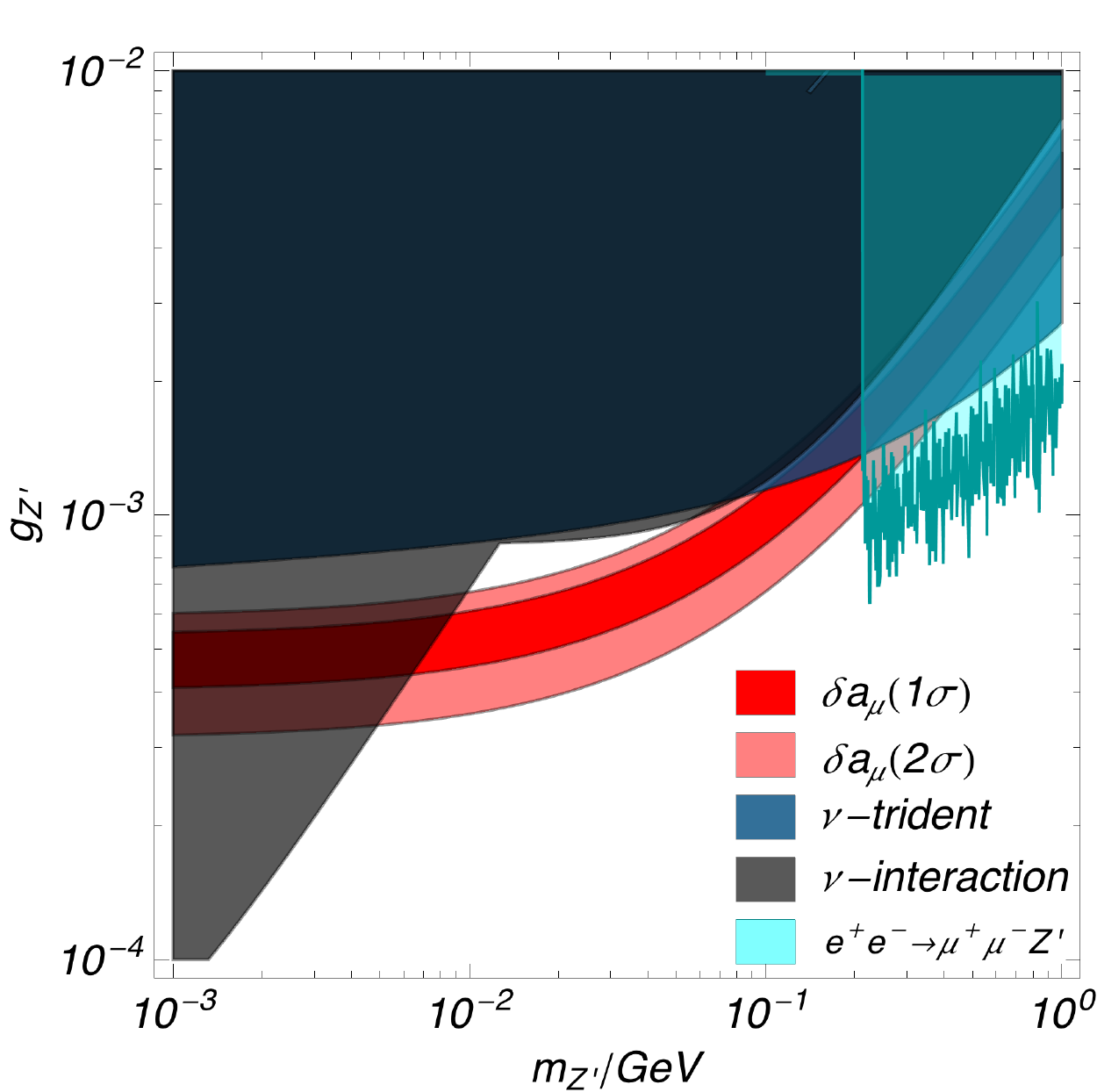}
 \end{minipage}
\end{center}
\caption{\sl \small The parameter region which explains the muon anomalous magnetic moment
within the $1\sigma$ (red) and the $2\sigma$ (pink) ranges.
The dark blue, gray and cyan shaded regions are excluded by the neutrino trident production experiments~\cite{Altmannshofer:2014pba}, neutrino-electron scattering experiments~\cite{Harnik:2012ni,Bilmis:2015lja} and $\mu^+\mu^-$ pair production searches associated with $Z'$  decaying into $\mu^+\mu^-$~\cite{TheBABAR:2016rlg} (Similar figure is given in \cite{Araki:2015mya}.), respectively. 
}
\label{fig:constraints}
\end{figure}

To evade those constraints, many models with lepton non-universal gauge charges 
have been discussed~\cite{Foot:1990mn,*Foot:1994vd,He:1991qd,Gninenko:2001hx,Baek:2001kca,Murakami:2001cs,
Ma:2001md,Pospelov:2008zw,Heeck:2011wj,Carone:2013uh,Heeck:2016xkh,Altmannshofer:2016brv}.
In particular, the model with the $L_\mu - L_\tau$ gauge interaction~\cite{Foot:1990mn,*Foot:1994vd,He:1991qd}
is the least constrained  due to its lack of  interactions with electrons,  electron-type neutrinos, and  quarks.%
\footnote{For extensions of the $L_\mu-L_\tau$ model coupling to the quark sector, see e.g.~\cite{Altmannshofer:2014cfa,Altmannshofer:2016jzy}
which also resolve the $B$-anomalies.}
Currently, the most severe constraint on this model comes from the neutrino trident production experiments~\cite{Geiregat:1990gz,Mishra:1991bv}.
As a consequence, the gauge boson mass, $m_{Z'}$, larger than about $400$\,MeV  favored by the muon anomalous magnetic moment~\cite{Altmannshofer:2014pba} is excluded. 
Some portion of the preferred parameter space has also been excluded by the searches
for the $\mu^+\mu^-$ pair production associated with a $Z'$ boson decaying into a $\mu^+\mu^-$ pair 
for $m_{Z'} \gtrsim 2\times m_\mu \simeq 211$\,MeV~\cite{TheBABAR:2016rlg} (see Fig.\,\ref{fig:constraints}).
Here, $m_\mu$ is the muon mass.

For  $m_{Z'} \lesssim 2\times m_\mu$, on the other hand, $Z'$ dominantly decays into neutrinos.
Thus, experimental tests of this model are even more difficult in this mass region.
In this paper, we discuss how the  $L_\mu - L_\tau$  model can be tested by looking for 
decays of the charged kaon into $Z'$, i.e. $K^+ \to \mu^+ +\nu_\mu + Z' (\to \nu\bar\nu)$,
where all the neutrinos in the final states are invisible.
As we will see, the favored parameter region for the muon anomalous magnetic moment can be 
tested by using a $10$ times larger number of stopped charged kaons and about a $100$ times better photon rejection rate
than the E949 experiment~\cite{Artamonov:2009sz,Artamonov:2014urb}.

The organization of the paper is as follows.
In section\,\ref{sec:model}, we summarize the $L_\mu-L_\tau$  model. 
In section\,\ref{sec:Kaon},  we discuss a testability of the model by using the decay of the stopped charged kaon associated with $Z'$.
Final section is devoted to our conclusions.

\section{$L_\mu-L_\tau$ model}\label{sec:model}
The lepton number symmetries for each flavor,  $L_i$ = ($L_e$, $L_\mu$, $L_\tau$), are not free from the quantum 
anomalies of the Standard Model gauge symmetry. 
Their differences,  $L_i - L_j$ ($i\neq j$), are, on the other hand, free from the quantum anomalies. 
Thus, they can be gauge symmetries without adding any extra charged fermions~\cite{Foot:1990mn,*Foot:1994vd,He:1991qd}.
Among them, the model with the $L_\mu - L_\tau$ gauge symmetry is particularly interesting 
where the gauge boson does not interact with electrons,  electron-type neutrinos, nor quarks. 

\begin{table}[t]
\caption{\sl\small
The charge assignment of the $L_\mu-L_\tau$  gauge symmetry.
Here, all the fermions are left-handed Weyl fermions.
The SM fields not in the table are not charged under the $L_\mu-L_\tau$ gauge symmetry.
}
\begin{center}
\begin{tabular}{|c|c|c|c|c|c|c|}
\hline
& $\ell_{\mu L} = (\nu_{\mu L}, \mu_L)^T$ & 
$\ell_{\tau L} = (\nu_{\tau L}, \tau_L)^T$  &
${\bar{\mu}_R} $ &
${\bar{\tau}_R} $ &
${\bar{N}_{\mu R}} $ &
 ${\bar{N}_{\tau R}}$ 
\\
\hline
$L_\mu-L_\tau$& $1$ & $-1$ & $-1$ & $1$ & $-1$ &$1$\\
\hline
\end{tabular}
\end{center}
\label{tab:charge}
\end{table}%

The gauge charge assignment of  the $L_\mu - L_\tau$ gauge symmetry is given in Table\,\ref{tab:charge}.
The $L_\mu - L_\tau$ gauge boson, $Z'$, couples to the SM fields  through the following Lagrangian,
\begin{eqnarray}
{\cal L}_{Z'}  &=& -\frac{1}{4}F_{Z'\mu\nu}F_{Z'}^{\mu\nu} + \frac{1}{2} m_{Z'}^2 {Z'}_\mu Z'^{\mu}
- g_{Z'}Z'_{\mu} j_{Z'}^\mu \ ,\\
j_{Z'}^{\mu} &=& \ell_{\mu L}^\dagger \bar\sigma^\mu \ell_{\mu L}
- \ell_{\tau L}^\dagger \bar\sigma^{\mu} \ell_{\tau L}
- \bar{\mu}^\dagger_R \bar\sigma^\mu\bar{\mu}_R 
+ \bar{\tau}^\dagger_R\bar\sigma^\mu\bar{\tau}_R 
- \bar{N}^\dagger_{\mu R} \bar\sigma^\mu\bar{N}_{\mu R} 
+ \bar{N}^\dagger_{\tau R}\bar\sigma^\mu\bar{N}_{\tau R}\ ,
\end{eqnarray}
where $F_{Z'}$ and $g_{Z'}$ denote the field strength
and the gauge coupling constant of $Z'$, respectively.
Here, we also assume spontaneous breaking of the $L_\mu-L_\tau$ gauge symmetry,
which leads to a mass of $Z'$, $m_{Z'}$.
In the gauge current $j_{Z'}^\mu$, $\ell_{\mu,\tau L}$, $\bar{\mu}_R$, and $\bar{\tau}_R$
denote the $\mu$ and $\tau$ doublet leptons, 
the $\mu$ and $\tau$ singlet leptons, respectively.
Here, we also include the right-handed neutrinos $\bar{N}_R$ to obtain SM neutrino masses  via the see-saw mechanism~\cite{Yanagida:1979as,Ramond:1979py}~\cite[see also][]{Minkowski:1977sc}.%
\footnote{See e.g. \cite{Harigaya:2013twa} for a model  which can reproduce the observed neutrino mixing angles.}
We also assume that ${\bar N}_R$'s obtain $L_\mu - L_\tau$ breaking scale masses, i.e. ${\cal O}(m_{Z'}/g_{Z'})$,
and hence, they do not play crucial roles in phenomenology in the following discussion.

\subsection{Muon anomalous magnetic moment}
Due to the non-vanishing charge of the muon, $Z'$ contributes to the muon anomalous magnetic moment at the one-loop level,
which is given by,
\begin{eqnarray}
\delta a_\mu = \frac{g_{Z'}^2}{8\pi^2}\int_0^1dx \frac{2 m_\mu^2 x^2(1-x)}{x^2 m_\mu^2 + (1-x) m_{Z'}^2}\ .
\end{eqnarray}
In Fig.\,\ref{fig:constraints}, we show the parameter region which explains the muon anomalous magnetic moment 
within the $1\sigma$ range as a red band.
The figure shows that the discrepancy of the muon anomalous magnetic moment 
can be resolved for $g_{Z'} \sim 5\times 10^{-4}$ for $m_{Z'} \ll m_\mu$ and 
$g_{Z'} \sim 5\times 10^{-3} (m_{Z'}/1\,{\rm GeV})$ for $m_{Z'} \gg m_\mu$.

In what follows, we summarize   constraints on the $L_\mu-L_\tau$ gauge theory.
Most of them are listed in Ref.\,\cite{Araki:2015mya} although we update  constraints from the neutrino interactions.

\subsection{Neutrino Trident Production}
The  results of the searches for the neutrino trident production, $\nu_\mu + N \to \nu_\mu + N + \mu^++\mu^-$, 
put severe constraints on the $L_\mu-L_\tau$ model where $N$ denotes a target nucleus~\cite{Altmannshofer:2014pba}.
In the SM, the neutrino trident production is mediated by the $W$ and $Z$ boson exchanges.
Similarly, the $Z'$ exchanges also contribute to the neutrino trident production.
The production cross section estimated from the CCFR experiment~\cite{Mishra:1991bv} 
is in good agreement with the SM prediction
\begin{eqnarray}
 \sigma_{CCFR}/ \sigma_{SM} = 0.82 \pm 0.28\ ,
 \end{eqnarray}
which leaves only a small room for the contribution from the $L_\mu-L_\tau$ interactions.%
\footnote{The CHARM-II collaboration also reported the trident event rate which is consistent with the SM prediction~\cite{Geiregat:1990gz},
from which we obtain a less stringent constraint on the $L_\mu-L_\tau$ interactions.}

In Fig.\,\ref{fig:constraints}, we show the 95\% CL exclusion limit on the $L_\mu-L_\tau$ model from the neutrino trident production.
Here, we use the equivalent photon approximation to 
estimate the production cross section according to~\cite{Altmannshofer:2014pba}.
The figure shows that the region favored by the muon anomalous magnetic moment 
is excluded for $m_{Z'} \gtrsim 400$\,MeV.

\subsection{Neutrino-Electron Interactions}
So far, we have assumed that the kinetic mixing between the photon and $Z'$,
\begin{eqnarray}
{\cal L}_{\rm mix} = \frac{1}{2}\epsilon F_{\mu \nu} F_{Z'}^{\mu \nu}\ ,
\end{eqnarray}
is vanishing, i.e. $\epsilon = 0$.
Such a kinetic mixing is, however, radiatively generated  even if we assume that it is vanishing at the tree-level.
At the one-loop level, the induced kinetic mixing parameter is given by,
\begin{eqnarray}
\label{eq:mix}
\epsilon = \frac{8}{3}\frac{eg_{Z'}}{16\pi^2} \log\frac{m_\tau}{m_\mu} \ ,
\end{eqnarray}
where $e$ is the QED coupling constant.%
\footnote{At the level of the QED, the kinetic mixing is forbidden by a discrete symmetry, 
$\mu \leftrightarrow \tau$, $F^{\mu\nu} \to F^{\mu\nu}$, and $F_{Z'}^{\mu\nu} \to- F_{Z'}^{\mu\nu}$
in the limit of $m_\mu = m_\tau$.
By the soft symmetry breaking $m_\mu\neq m_\tau$, radiative corrections generate a finite kinetic mixing. 
}

Once $Z'$ has a kinetic mixing to the photon, $Z'$ obtains couplings to the QED current $j^\mu_{\rm QED}$
\begin{eqnarray}
{\cal L} \simeq -\epsilon e Z'_\mu j^\mu_{\rm QED}\ ,
\end{eqnarray}
after eliminating the kinetic mixing term by shifting the photon fields.
Thus, the $L_\mu-L_\tau$ model can be 
further tested through the interactions of $Z'$ with the electrons and quarks.%
\footnote{The interactions of $Z'$ with  electron-type neutrinos are still suppressed
since the effects of the kinetic mixing to $Z$ boson are suppressed by $m_Z'^2/m_Z^2$.}
In particular, the neutrino-electron scattering experiments put severe constraints~\cite{Harnik:2012ni,Bilmis:2015lja}.

In Fig.\,\ref{fig:constraints}, we show the limits from the neutrino-electron scattering experiments at the 90\% CL exclusion limit,
which are translated from the ones obtained in \cite{Bilmis:2015lja}.
It should be noted that the $L_\mu-L_\tau$ model cannot be constrained by the experiments using
$\nu_e$ nor $\bar{\nu}_e$ unless they oscillate into other flavors.
Thus, the TEXONO experiments~\cite{Deniz:2009mu,Li:2002pn,*Wong:2006nx,Chen:2014dsa} 
which put the most stringent limits on the flavor universal gauge interactions 
do not constrain the $L_\mu-L_\tau$ model.
As a consequence, we find that the primary constraints come from the CHARM-II experiment~\cite{Vilain:1993kd,Vilain:1994qy}
for $m_{Z'} \gtrsim 200$\,MeV which uses the $\nu_\mu$ and $\bar{\nu}_\mu$ beams, 
and from the BOREXINO experiment~\cite{Bellini:2011rx} for $m_{Z'}\lesssim 200$\,MeV
where about a half of $^7$Be solar neutrinos oscillate into other neutrinos.

\subsection{$e^++e^-$ Collider Experiment }
The BaBar experiments put constraints on the $\mu^+\mu^-$ pair production associated with $Z'$ where $Z'$ decays into $\mu^+\mu^-$~\cite{TheBABAR:2016rlg}.
In Fig.\,\ref{fig:constraints}, we show the constraints at the 90\% CL translated from~\cite{TheBABAR:2016rlg} (the cyan shaded region). 

Let us also comment on the beam dump experiments utilizing the electron or the proton beams
which put severe limits on the sub-GeV dark photon models with a kinetic mixing to the photon
of $\epsilon \sim 10^{-(5-6)}$~\cite{Essig:2010gu}.
In the $L_\mu-L_\tau$ model, however, $Z'$ in the $L_\mu-L_\tau$ model immediately decays into neutrinos, 
and hence, they do not lead to stringent limits, despite the non-vanishing kinetic mixing as in Eq.\,(\ref{eq:mix}).

\subsection{Other Constraints}
Before closing this section, let us discuss other constraints which are not shown in Fig.\,\ref{fig:constraints}.
First, in order not to spoil the success of the Big-Bang Nucleosynthesis (BBN), 
additional contributions to the effective number of relativistic species at around the BBN temperature, $N_{\rm eff}$,
are limited to be ${\mit \Delta}N_{\rm eff} \lesssim 1$~\cite{Mangano:2011ar,Steigman:2012ve}.
This constraint puts a lower limit on the $Z'$ mass, $m_{Z'} \gtrsim 5$\,MeV~\cite{Kamada:2015era}.

When the $Z'$ mass is around or below the typical core temperature of the supernovae, $T\sim 30$\,MeV,
$Z'$  can be produced inside the cores of the supernovae.
The presence of $Z'$ in the supernova cores can affect the diffusion times of the neutrinos which 
should be around 10\,s estimated from the observed duration of the neutrino burst of SN1987A~\cite{Hirata:1987hu,Bionta:1987qt}.
According to Ref.~\cite{Kamada:2015era}, this constraints exclude the parameter region favored by the muon anomalous magnetic 
moment with $m_{Z'} \lesssim 30$--$50$\,MeV.

For $m_{Z'} > 2m_\mu$, the SM $Z$ boson decays into a pair of $\mu^+\mu^-$ associated with 
the $Z'$ production which subsequently decays into $\mu^+\mu^-$~\cite{Heeck:2011wj,Harigaya:2013twa,Elahi:2015vzh}.
However, the resultant limits from the LHC experiments are  less stringent from the ones shown in Fig.\,\ref{fig:constraints}.

In summary, the $L_\mu-L_\tau$ interactions can successfully resolve the discrepancy of the muon anomalous magnetic moment 
for ${\cal O}(1)\,{\rm MeV} \lesssim m_{Z'} \lesssim 400$\,MeV and $g_{Z'} \sim 10^{-3}$, while evading all the experimental constraints.
In the following, we discuss how we can test the remaining parameter region by using the rare kaon decay.%
\footnote{In~\cite{Gninenko:2014pea}, it is proposed to look for $Z'$ in the $\mu+Z \to \mu + Z + Z'(\to \nu\bar\nu)$ reaction, 
which reaches down to $g_{Z'} = {\cal O}(10^{-5})$ by using ${\cal O}(10^{12})$ incident muons at the energy $E_\mu = 150$\,GeV. }

\section{Rare Kaon Decay}\label{sec:Kaon}
For $m_{Z'} < 2\times m_\mu$,  $Z'$ decays into neutrinos 
and hence, the rare charged kaon decay mode,  $K^+ \to \mu^+  + \nu_\mu + Z'$,
results in $K^+ \to \mu +{\rm invisible}$.
This mode can be distinguished from the main mode of the charged kaon, 
$K^+\to \mu ^+ + \nu_\mu$, since the main mode emits the monochromatic muon with a momentum, $p_\mu = 236$\,MeV,
while the muon in the $K^+ \to \mu^+  + \nu_\mu + Z'$ mode possesses a continuous spectrum.%
\footnote{The branching ratio of the irreducible background in the SM, $K^+ \to \mu^+ + \nu_\mu + \nu + \bar{\nu}$,
 is predicted to be ${\cal O}(10^{-16})$~\cite{Gorbunov:2016tbk}.} 
(See \cite{Carlson:2012pc,Beranek:2012ey,Laha:2013xua} for earlier works to utilize $K^+ \to \mu +{\rm invisible}$ mode
to put constraints on light particles coupling to the muon.)

\begin{figure}[tbp]
	\centering
		\begin{minipage}{.46\linewidth}
  \includegraphics[width=\linewidth]{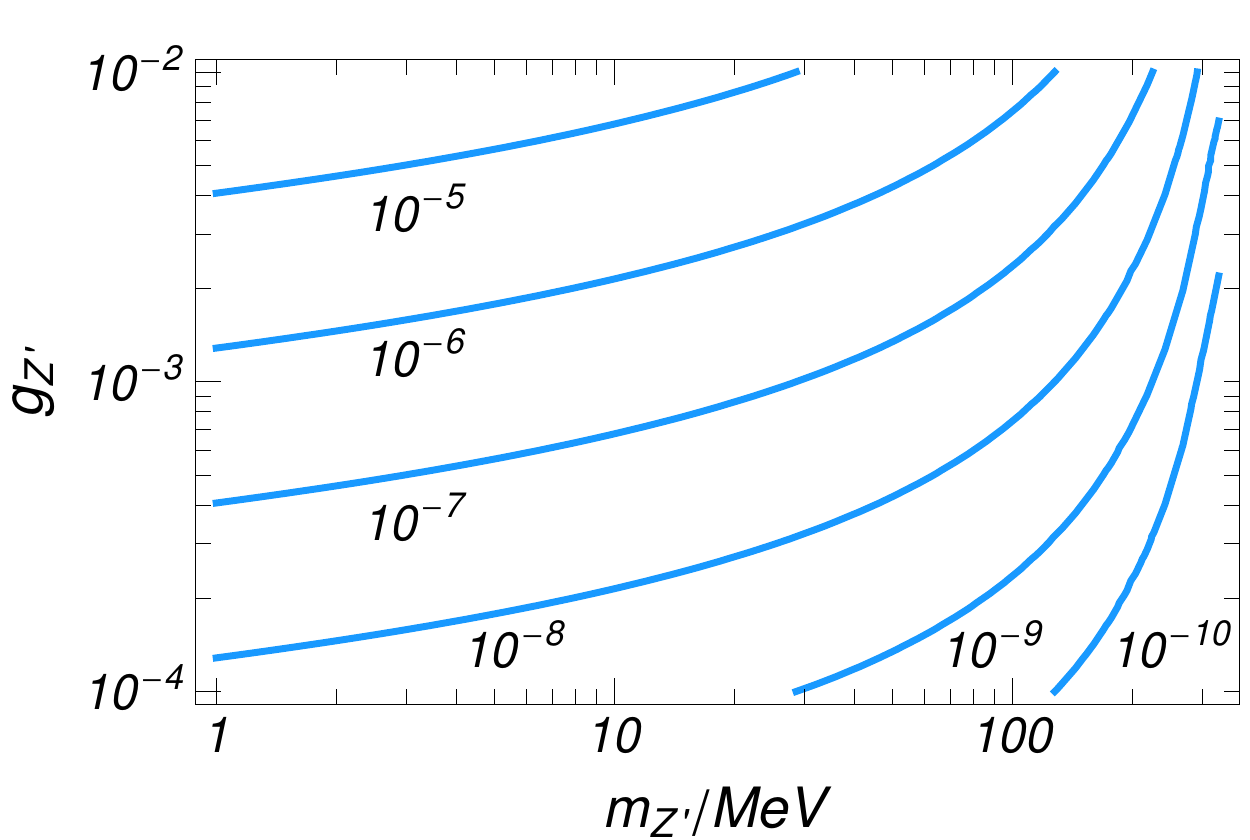}
 \end{minipage}
 \hspace{1cm}
 \begin{minipage}{.46\linewidth}
  \includegraphics[width=\linewidth]{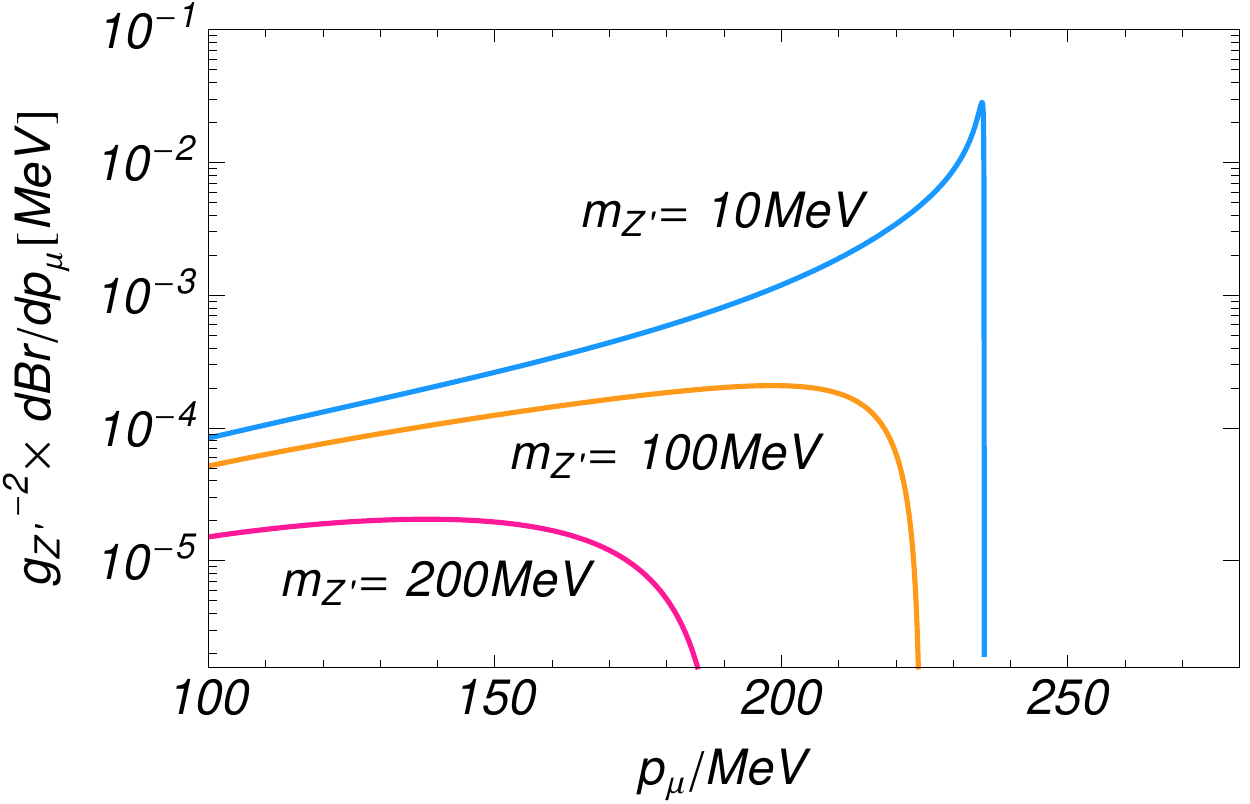}
 \end{minipage}
	\caption{Left) The branching ratio of $K^+$ into $K^+\to \mu^+ + \nu_\mu + Z'$.
	Right) The differential branching ratio  normalized by $g_{Z'}^{-2}$ for given $Z'$ masses.
		}
	\label{fig:BR}
\end{figure}

The amplitude of the decay mode, $K^+ \to \mu^+ + \nu_\mu+Z'$, is given by 
\begin{eqnarray}
\label{eq:MLmuLtau2}
{\cal M}  &=&- i 2  f_K V_{us} G_F g_{Z'} \, \nonumber\\ 
&&\times 
\varepsilon^{\nu*}(Z')
\left\{
\bar{u}(\nu_\mu) 
P_R  \slashchar{\ell}
\frac{2\,p_{\nu} + \slashchar{q}\gamma_\nu}{2\,p\cdot q + q^2 } 
v(\mu^+)
-
\bar{u}(\nu_\mu)P_R
\frac{(2\,{k}_{\nu} + \gamma_\nu \slashchar{q})}{ 2 k\cdot q + q^2 } 
 \slashchar{\ell} v(\mu^+)
\right\} \ ,  
\end{eqnarray}
where $G_F$  is the Fermi constant, $V_{us}\simeq 0.23$  the CKM angle,
and $f_K$  the kaon decay constant.
The four momenta of $\mu^+$, $\nu_\mu$, $Z'$ and $K^+$ are denoted by $p$, $k$, $q$ and $\ell$, respectively. 
The wave functions of $\mu^+$, $\nu_\mu$ and $Z'$ are denoted by $v(\mu^+)$, $u(\nu_\mu)$, and $\varepsilon(Z')$, respectively.

The spin-summed squared amplitude is then given by
\begin{eqnarray}
&& \sum_{\mathrm{spin}} |\overline{\cal M}|^2 = 4 f_K^2 |V_{us}|^2 G_F^2 g_{Z'}^2\times
\frac{ 2 m_\mu^2} {m_{12}^4 (m_{23}^2-m_\mu^2)^2}\times
\nonumber \\ 
&&
\bigg({M_{Z'}^2} 
\left({M_K^2} \left(-{m_{12}^4}-{m_{23}^4}\right)
+{m_{\mu}^2} \left(({m_{12}^2}+{m_{23}^2})^2+2 {m_{23}^2}  {M_K^2}\right) 
-{m_{\mu}^4}\left(2 ({m_{12}^2}+{m_{23}^2})+{M_K^2}\right)
+{m_{\mu}^6}\right) 
\nonumber \\ 
&&+{m_{12}^2} \left({m_{23}^2} \left(2{M_K^4}-2 {M_K^2}
   ({m_{12}^2}+{m_{23}^2})+({m_{12}^2}+{m_{23}^2})^2
   \right)
-{m_{\mu}^2}
   \left(({m_{12}^2}+{m_{23}^2})^2+2 {M_K^4}\right) \right.
\nonumber   \\
&&
\left.  +{m_{\mu}}^4 
  \left(2  {m_{12}^2}+{m_{23}^2}+2 {M_K^2}\right)
   -{m_{\mu}^6}\right)\bigg) \ .
\end{eqnarray}
Here, $M_K \simeq  493.7$\,MeV is the charged kaon mass, and the Dalitz parameters $m_{12}^2$ and $m_{23}^2$ are defined by,
\begin{eqnarray}
m_{12}^2 = (k+q)^2\ , \quad m_{23}^2 = (p+q)^2\ ,
\end{eqnarray}
respectively.
The pre-factor of the squared amplitude can be read off from the decay width of main decay mode,
\begin{eqnarray}
 f_K^2 |V_{us}|^2 G_F^2  
 \simeq \frac{4\pi M_K^3}{m_\mu^2(M_K^2-m_\mu^2)^2} \Gamma(K^+\to \mu^+ + \nu_\mu) \ , \\
 \simeq \frac{4\pi M_K^3}{m_\mu^2(M_K^2-m_\mu^2)^2} \,Br(K^+\to \mu^+ + \nu_\mu) \,\Gamma_K\ ,
\end{eqnarray}
where the branching fraction of the main mode and the total decay width $\Gamma_K$ 
are given by\,\cite{Olive:2016xmw},
\begin{eqnarray}
Br(K^+\to \mu^+ + \nu_\mu) = 0.6356 \pm 0.0011\ , \quad 
\Gamma_K^{-1} =  \left(1.2380\pm 0.0020\right) \times 10^{-8}\, {\rm s} \ ,
\end{eqnarray}
respectively.
Altogether, the decay width into $Z'$ is given by,
\begin{eqnarray}
\label{eq:width}
\Gamma(K^+ \to \mu^++\nu_\nu+ Z')  = \frac{1}{32  M_K^3}\frac{1}{(2\pi)^3}\int dm_{12}^2 \, dm_{23}^2 \,|\overline {\cal M}|^2\ . 
\end{eqnarray}
The ranges of the Dalitz parameters, $m_{12}^2$ and $m_{23}^2$, are summarized in  appendix\,\ref{sec:Dalitz}.
The differential decay width for the muon momentum $p_\mu$ is also given by, 
\begin{eqnarray}
\label{eq:dwidth}
\frac{d}{d{p}_\mu}\Gamma(K^+ \to \mu^++\nu_\nu+ Z')  = \frac{1}{32  M_K^3}\frac{1}{(2\pi)^3}
\frac{2 M_K p_\mu}{\sqrt{m_\mu^2 + p_\mu^2}}
\int dm_{23}^2 \,|\overline {\cal M}|^2\ , 
\end{eqnarray}
where $m_{12}^2$ is related to $p_\mu$ via
\begin{eqnarray}
\label{eq:m12}
 m_{12}^2 = M_K^2 + m_\mu^2   - 2 M_K \sqrt{m_\mu^2 + p_\mu^2}\ ,
\end{eqnarray}
in the charged kaon rest frame.
The range of the muon momentum $p_\mu$ in Eq.\,(\ref{eq:dwidth}) is given by
\begin{eqnarray}
 0 < p_\mu <  \frac{1}{2 M_K} \left((M_{K}^2 -(m_\mu+ m_{Z'})^2)
 (M_K^2 -(m_\mu-m_{Z'})^2)\right)^{1/2} \ ,
\end{eqnarray}
which can be read off from  Eqs.\,(\ref{eq:m12}), (\ref{eq:m12min}) and (\ref{eq:m12max}).

In Fig.\,\ref{fig:BR}, we show the branching ratio and the muon momentum spectrum
in the $L_\mu-L_\tau $ model.
The muon track searches in the momentum range $128$\,MeV to $176$\,MeV 
from the decays of the stopped charged kaons at the LBL Bevatron
put the upper limits the branching ratio into $K^+ \to \mu^+ + {\rm invisible}$ of $6\times 10^{-6}$~\cite{Pang:1989ut}.
By compared with the predicted branching ratio in Fig.\,\ref{fig:BR}, we find that this limit does not exclude
the parameter region which explains the muon anomalous magnetic moment.

To put a more stringent constraint, we consider the search for  heavy neutrinos in
the decay of the stopped charged kaon by the E949 experiment~\cite{Artamonov:2014urb}.
There, single muon tracks in the momentum range $140$\,MeV to $200$\,MeV
were searched for with a total exposure of $N_K^{\rm E949} = 1.70\times 10^{12}$ stopped kaons.%
\footnote{Typically, $1.6\times 10^6$ kaons per second enter the stopping target.}
The main background to the signal comes from 
the radiative decay mode, $K^+ \to \mu^+ + \nu_\mu + \gamma$, 
where the photon is misidentified.%
\footnote{The decay modes involving charged pions are effectively vetoed by the Range-Momentum cut~\cite{Artamonov:2014urb}.}
In the E949 experiment, a photon rejection rate is around $\epsilon^{\rm E949}_\gamma \sim10^{-3}$ after tight photon veto cuts.
By comparing with the branching ratio of the radiative decay mode~\cite{Agashe:2014kda},%
\footnote{In this momentum range, the so-called internal bremsstrahlung dominates the radiative decay.}
\begin{eqnarray}
BR(K^+ \to \mu^+ + \nu_\mu + \gamma, 140\,{\rm MeV}<p_\mu<200\,{\rm MeV}) = (1.4 \pm 0.2)\times 10^{-3}\ ,
\end{eqnarray}
we expect that the results of the E949 experiment can put the limits of the branching ratio into $Z'$ of ${\cal O}(10^{-6})$.
In fact, the E949 collaboration recently reported a limit on a branching ratio, $BR(K^+ \to \mu+  \nu\nu\bar{\nu}) \lesssim 2.4\times 10^{-6}$
at the 90\% CL~\cite{Artamonov:2016wby}, by assuming non-standard dimension six neutrino interactions~\cite{Bardin:1970wq}.

\begin{figure}[tbp]
	\centering
		\begin{minipage}{.46\linewidth}
  \includegraphics[width=\linewidth]{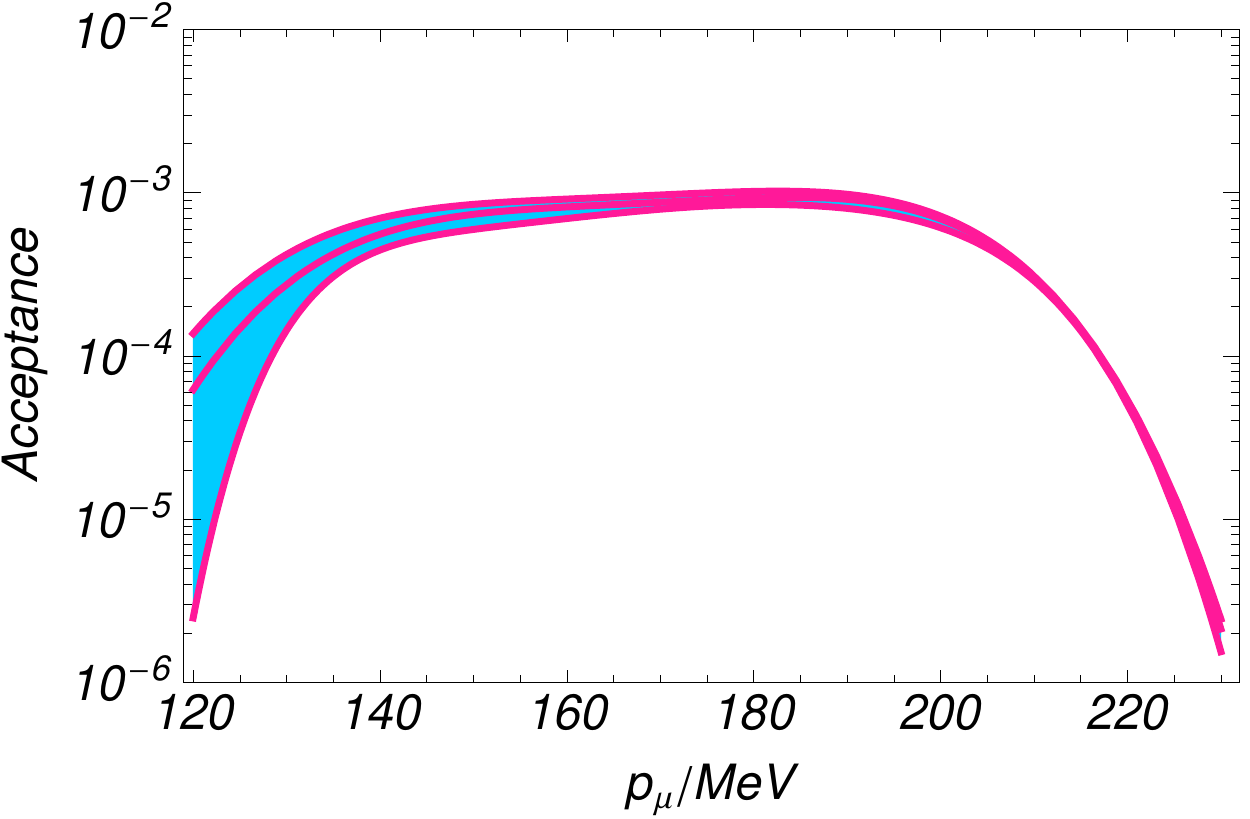}
 \end{minipage}
 \hspace{1cm}
 \begin{minipage}{.46\linewidth}
  \includegraphics[width=\linewidth]{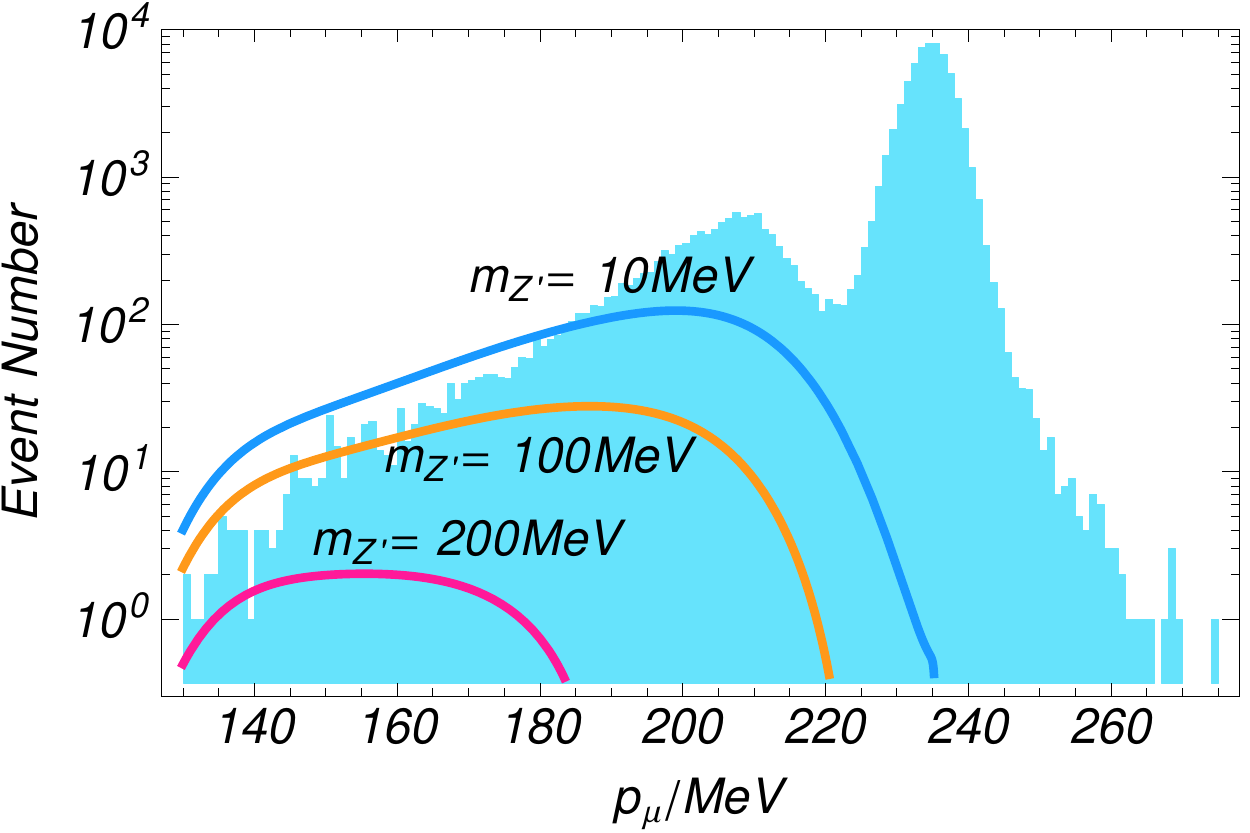}
 \end{minipage}
	\caption{Left) The single-muon acceptance in the E949 experiment after all the cuts are applied~\cite{Artamonov:2014urb}. 
	The band corresponds to the $1\sigma$ uncertainties of the single-muon acceptance. 
	Right) The signal spectrum expected at the E949 experiment where the muon acceptance is assumed to be 
	the lower one in the left panel.
	We also show the observed event numbers at the E949 experiment after tight photon veto cuts~\cite{Artamonov:2014urb}.
		}
	\label{fig:spectrum}
\end{figure}

In the left panel of Fig.\,\ref{fig:spectrum}, we show a single-muon acceptance after all the cuts are applied~\cite{Artamonov:2014urb}.
The figure indicates that the acceptance is around $10^{-3}$ for $140\,{\rm MeV}<p_\mu<200\,{\rm MeV}$.
The band corresponds to the $1\sigma$ error of the muon acceptance for a given muon momentum.
In the right panel, we also show the expected muon spectrum in the $L_\mu-L_\tau$ model 
 at the  E949 experiment for $g_{Z'} = 10^{-2}$, i.e.
 \begin{eqnarray}
\label{eq:dwidth}
\frac{dN_{L_\mu-L_\tau}}{dp_\mu} = \frac{1}{\Gamma_K}\frac{d}{d{p}_\mu}\Gamma(K^+ \to \mu^++\nu_\nu+ Z')   \times N_{K}^{\rm E949} 
\times {\mbox{(single muon acceptance)}}\ .
\end{eqnarray}
Here, we take the lower muon acceptance in the left panel to make our analysis conservative. 
The histogram in the figure shows the observed event numbers at the E949 experiment after the tight photon veto cuts in
\cite{Artamonov:2014urb}.

To estimate conservative exclusion limits on the gauge coupling constant for a given $m_{Z'}$,
we combine the event numbers in the momentum bins from $p_\mu =130$\,MeV to $p_{\mu}= 160$\,MeV
into one bin.
Then, we use a test statistic,
\begin{eqnarray}
\label{eq:likelihood}
 -2 \ln {\cal L} =  2\left( N_s-N_{\rm obs} + N_{\rm obs} \ln \frac{N_{\rm obs}} {N_s} \right)\ ,
\end{eqnarray}
where $N_{\rm obs}$ and $N_s$ are the  observed and the  expected signal event numbers 
in the combined bin.
It should be noted that a more stringent limit can be obtained if the spectrum
of the background muons including  detector responses is accurately predicted
so that a multi-bin analysis can be performed.

\begin{figure}[t]
\begin{center}
\begin{minipage}{.5\linewidth}
  \includegraphics[width=\linewidth]{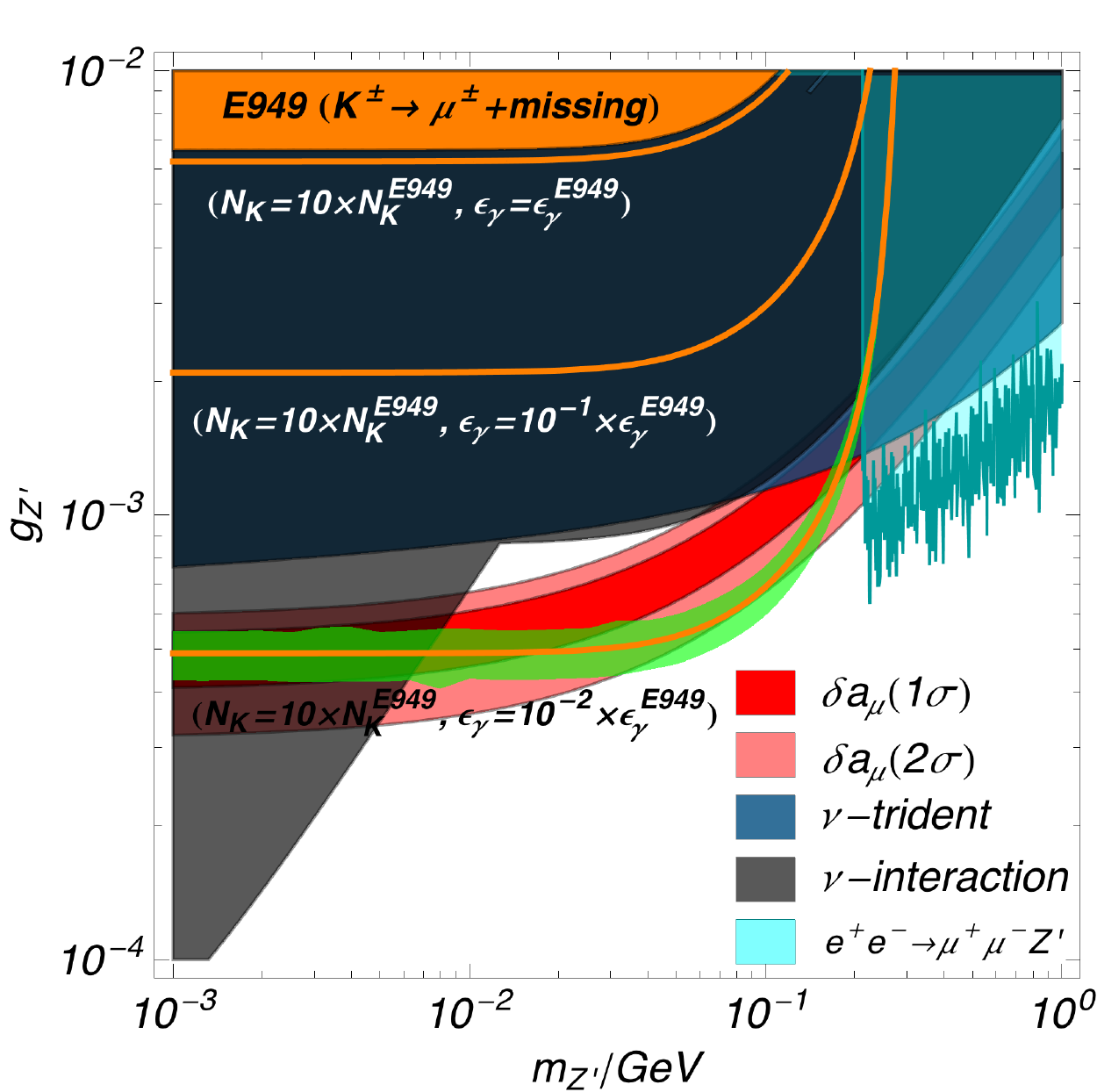}
 \end{minipage}
\end{center}
\caption{\sl \small The 95\,\% CL limit from the E949 experiment (the orange shaded region)
overlaid on Fig.\,{\ref{fig:constraints}}.
We also show possible the improved sensitivities by assuming several setups as indicated in the figure.
The green band around the expected exclusion limit for $N_K =10\times N_K^{\rm E949}$
and $\epsilon_\gamma = 10^{-2} \times {\epsilon}_\gamma^{\rm E949}$ 
corresponds to the $2\sigma$ statistical fluctuation of the expected exclusion limit. 
}
\label{fig:final}
\end{figure}

In Fig.\,\ref{fig:final}, we show the 95\,\% CL limit from the E949 experiment as the orange shaded region.
Here, we define the 95\,\% CL limit by $-2 \ln{\cal L} > \chi^2_{95}= 3.842$ assuming that $N_s \gg 1$.
The figure shows that the results of the E949 experiment exclude the parameter region
corresponding to the  branching ratio into $Z'$ of ${\cal O}(10^{-6})$ as expected.
It should be noted that the exclusion limit is insensitive to $m_{Z'}$ in a lighter $Z'$ region 
while the predicted branching ratio in Fig.\,\ref{fig:BR} is larger for a lighter $Z'$ for a given $g_{Z'}$.
The insensitivity of the constraints on $m_{Z'}$ can be understood from the spectrum shapes of the signal in Fig.\,\ref{fig:spectrum},
which shows the number of events in the momentum bins from $p_\mu =130$\,MeV to $p_{\mu}= 160$\,MeV
is less sensitive to $m_{Z'}$ in the light $Z'$ region.

Now, let us discuss how the constraint can be improved by assuming several experimental setups.
First, let us consider a $10$ times larger exposure,  $N_K = 10\times N_K^{E949}$,
while considering the same muon acceptance and the same photon rejection rate compared with the E949 experiment.
In the E949 experiment, $5\times 10^5$ kaons are yielded from every $10^{12}$ p.o.t.~\cite{Artamonov:2009sz}.  
Thus,  $N_K =  10\times N_K^{E949}$ is easily achieved if we assume the proton beam of ${\cal O}(10^{20})$ p.o.t. 
which corresponds to the assumption for the SHiP experiment~\cite{Anelli:2015pba}
and the TREK experiment~\cite{TREK:2010,*Lu:2016onh}. 
As the figure shows, however, the mere large statistics does not improve the expected exclusion limit very much.

Next, let us assume $10$ and $100$ times better photon rejection rates compared with the E949 experiments
while assuming the same muon acceptance.
For feasibility of those high rejection rates, see e.g.~\cite{Ramberg:2004en}.
The figure shows that with a $100$ times better photon rejection rate with $N_K = 10\times N_{K}^{\rm E949}$,
a significant portion of the parameter region favored by the muon anomalous magnetic moment can be tested.%
\footnote{Even for a $100$ times better rejection rate, we find that $N_s$ at the 95\% CL limit corresponds to 
$N_s \simeq 8$, which is large enough to approximate the distribution of $-2\ln{\cal L}$ by the $\chi^2$ distribution. }  
In the figure, we also show the $2\sigma $ statistical fluctuation of the expected exclusion limit
for  a 100 times better photon rejection rate (see  appendix\,\ref{sec:fluctuation}).

Before closing this section, let us comment the kaon experiments in which the kaons 
decay in-flight as in the NA62 experiment~\cite{Goudzovski:2012gh} and the SHiP experiment~\cite{Anelli:2015pba}.
In those types of experiments, the kaon beams are contaminated by the muons.
Those muon contaminations contribute to background events for the searches of single muons in $K^+ \to \mu +{\rm invisible}$.
Thus, in the in-flight kaon decay experiments, it is important to develop a veto system 
to reject the muons in the beams while accepting the muons from the decays of the kaons.
For the experiments with stopped kaons such as the TREK experiment~\cite{TREK:2010,*Lu:2016onh}, 
on the other hand, the muon contaminations in the beam do not contribute to background events.

\section{Conclusions}
The model with the $L_\mu -L_\tau$ gauge symmetry is the least constrained model 
as a resolution to the discrepancy of the muon anomalous magnetic moment between the 
theoretical prediction and the experimental result.
There, the gauge boson of the $L_\mu-L_\tau$ gauge symmetry, $Z'$, has highly suppressed interactions 
with electrons,  electron-type neutrinos, and quarks.
The experimental test of the model is particularly challenging when the $Z'$ mass is lower than the muon pair threshold. 

In this paper, we discussed how well the  $L_\mu - L_\tau$  model could be tested  by using the decay of the charged kaon associated with $Z'$.
In particular, we consider the constraints by looking for a single muon track from the decay of the stopped charged kaons
as in the E949 experiment~\cite{Artamonov:2014urb,Artamonov:2016wby}.
 According to our conservative estimation, we find that the large portion of the  favored parameter region for the muon anomalous magnetic moment can be 
tested by using a $10$ times larger number of stopped charged kaons and about a $100$ times better photon rejection rate.

It should be stressed that the constraints derived in this paper are based on very conservative assumptions.
For example, the charged kaon exposure much larger than $N_K = 10\times N_K^{E949}$ can be achieved 
with a beam intensity assumed  for the TREK experiment~\cite{TREK:2010,*Lu:2016onh}.
With such large statistics, it is possible to optimize the analysis by using only low muon momentum bins.
There the photons in the background events tend to have higher energies, and are vetoed more efficiently~\cite{Ramberg:2004en}.
Furthermore, with large statistics, it is also possible to apply tighter photon veto cuts while obtaining a single-muon 
acceptance after the photon veto cuts in a data-driven manner.
With a reliable estimation of the muon acceptance including the effects of detector responses, a multi-bin analysis  enhances the sensitivity due to the difference of the $p_\mu$ spectrum shapes between 
the signal and the background.
With those optimizations, future experiments such as the TREK (E06) experiment proposed at J-PARC~\cite{TREK:2010,*Lu:2016onh}
are expected to improve the exclusion limit greatly.%
\footnote{
As discussed in \cite{Altmannshofer:2014pba}, 
it is also possible to test the favored parameter region for the muon anomalous magnetic moment 
by the neutrino trident production at future neutrino facilities, such as LBNE, by assuming 18-ton Argon near detectors and the proton beam of ${\cal O}(10^{20})$ p.o.t.,
which also requires more detailed numerical studies.}

\begin{acknowledgments}
MI thank Satoshi Shirai for useful discussion.
This work is supported in part by Grants-in-Aid for Scientific Research from the Ministry of Education, Culture, Sports, Science, and Technology (MEXT), Japan, No. 25105011 and No. 15H05889 (M. I.); Grant-in-Aid No. 26287039 (M. I.) from the Japan Society for the Promotion of Science (JSPS); and by the World Premier International Research Center Initiative (WPI), MEXT, Japan.
\end{acknowledgments}

\appendix
\section{The Dalitz parameters}
\label{sec:Dalitz}
In this section, we summarize the range of the Dalitz parameters used in the decay width in Eq.\,(\ref{eq:width}).
The first Dalitz parameter $m_{12}^2$ is the invariant mass of the $\nu_\mu$ and $Z'$ system which ranges between
\begin{eqnarray}
\label{eq:m12min}
m_{12}^2|_{\rm min}  &=& m_{Z'}^2 \ ,\\
\label{eq:m12max}
m_{12}^2|_{\rm max}  &=& (M_K-m_\mu)^2 \ .
\end{eqnarray}
The minimum corresponds to the final state with the vanishing $\nu_\mu$ momentum in the rest frame of the charged kaon,
while the maximum corresponds to the one with the vanishing $p_\mu$ in the rest frame of the charged kaon.

For a given $m_{12}^2$, the minimum and the maximum of $m_{23}^2$ are given by 
\begin{eqnarray}
m_{23}^2|_{\rm min} & = & (E_2^* + E_3^*)^2 - 
\left(\sqrt{E_2^{*2} - m_{Z'}^2} + 
\sqrt{E_3^{*2} - m_{\mu}^2}  \right)^2 \ , \nonumber\\
m_{23}^2|_{\rm max} & = & (E_2^* + E_3^*)^2 - 
\left(\sqrt{E_2^{*2} - m_{Z'}^2} - 
\sqrt{E_3^{*2} - m_{\mu}^2}  \right)^2 \ ,
\end{eqnarray}
where
\begin{eqnarray}
E_2^* &=& \frac{m_{12}^2 + m_{Z'}^2}{2 m_{12}}\ , \\
E_3^* &=& \frac{M_{K}^2 - m_{12}^2 - m_\mu^2}{2 m_{12}}\ ,
\end{eqnarray}
(see e.g.  \cite{Olive:2016xmw}).

\section{Statistical Fluctuation of the Expected Exclusion Limits}
\label{sec:fluctuation}
\begin{figure}[tbp]
	\centering
		\begin{minipage}{.46\linewidth}
  \includegraphics[width=\linewidth]{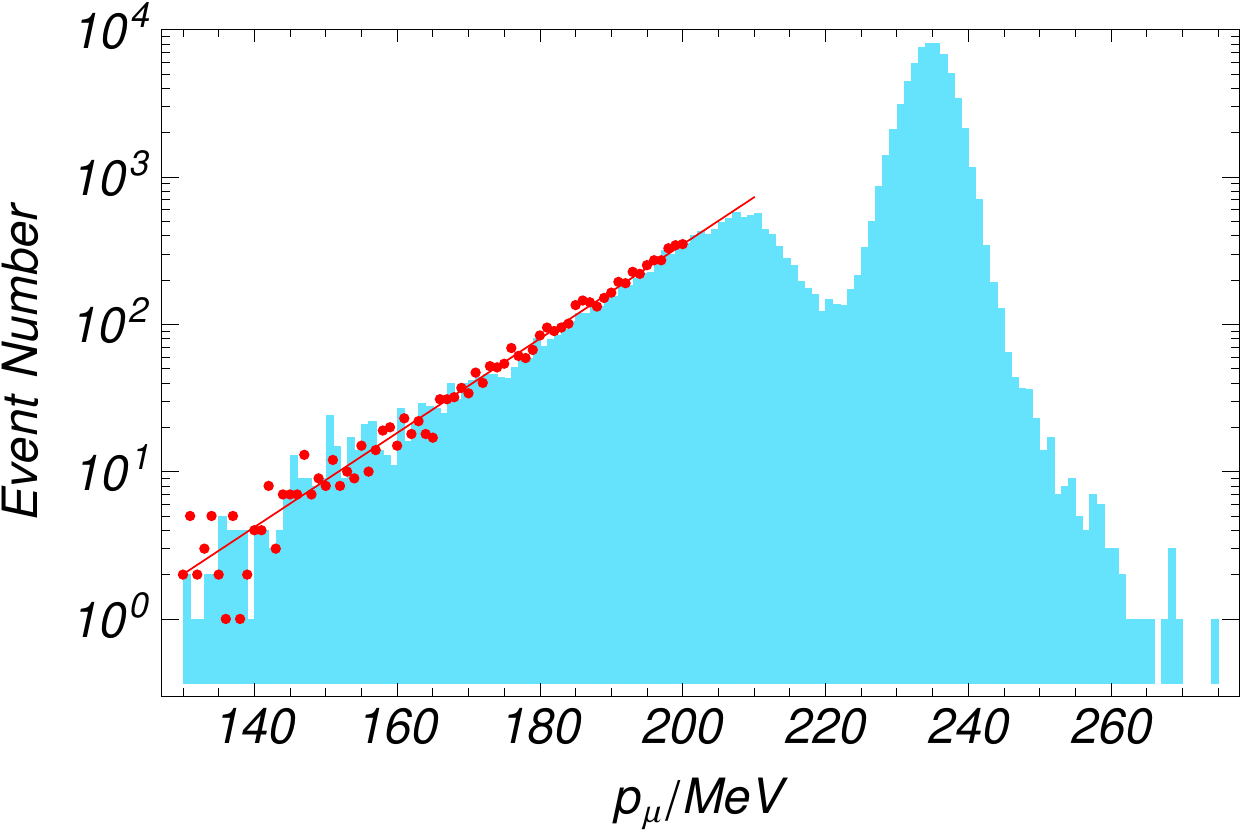}
 \end{minipage}
 \hspace{1cm}
 \begin{minipage}{.46\linewidth}
  \includegraphics[width=\linewidth]{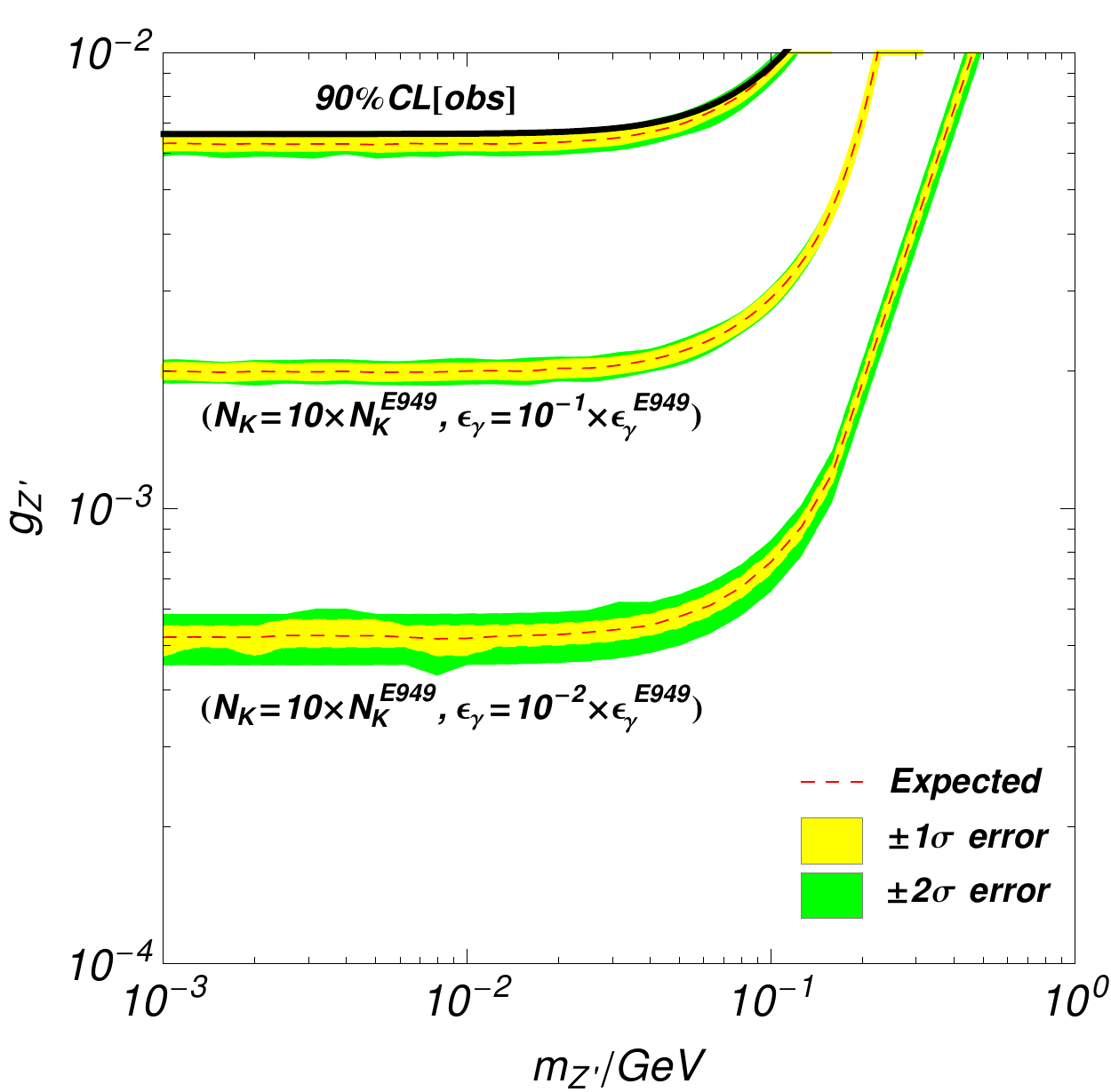}
 \end{minipage}
	\caption{Left) The fitted spectrum used to generate mock sample data sets (red line).
	One example of the mock data is also shown by red points.
	Right) The statistical fluctuations of the expected exclusion limits at 95\% CL for
	various assumptions.
	The yellow shaded bands and the green bands correspond to the 1$\sigma$ and the 2$\sigma$
	statistical fluctuations, respectively.
	The dashed lines show the mean expected exclusion limits.
	The black line is the observed exclusion limit obtained in the main text by using the E949 data.
	}
	\label{fig:Mock}
\end{figure}

To estimate the statistical fluctuation of the expected exclusion limit, 
we generate mock data samples of the background events.
For that purpose, we use the fitted spectrum shown in Fig.\,\ref{fig:Mock} as a background model 
since it is difficult to predict the background spectrum precisely including the detector responses.
To generate the mock data for various assumptions, we simply scale the fitted spectrum by factors of $N_K/N_K^{\rm E949}$
and $\epsilon_{\gamma}/\epsilon_{\gamma}^{\rm E949}$.

In the right panel of Fig.\,\ref{fig:Mock}, we show the statistical fluctuations of the expected exclusion limits at the 95\% CL for
various assumptions.
Here, we use the same statistical analysis used in the main text (see discussion around Eq.\,(\ref{eq:likelihood})).
For the original setup of the E949 experiment, the mean expected limit is slightly more stringent than 
the observed one in Fig.\,\ref{fig:final}.
This discrepancy indicates that our fitted spectrum slightly deviates from the true background model,
although the discrepancy is within the $2\sigma $ range.
The statistical fluctuations around the mean expected limits are, on the other hand, 
not expected to be  very sensitive to the shape of the background spectrum. 
In Fig.\,\ref{fig:final} in the main text, we show the fluctuation for $N_K = 10\times N_K^{E949}$ and $\epsilon_\gamma = 10^{-2}\times \epsilon_{\gamma}^{\rm E949}$
in Fig.\,\ref{fig:Mock} around the expected exclusion limit in the analysis of the main text.

\bibliography{draft_revised}

\end{document}